\documentstyle[preprint,aps]{revtex}
\begin{document}

\draft

\title{Differential Regularization of Chern-Simons-Maxwell Spinor 
and Scalar Electrodynamics }
\author{M. Chaichian${}^{a,b}$, W. F. Chen${}^{b}$\renewcommand{\thefootnote}
{\dagger}\footnote{\small ICSC-World Laboratory, Switzerland} and 
 H.C. Lee${}^c$}
\address{${}^a$ High Energy Physics Division, Department of Physics,
Univeristy of Helsinki\\
${}^b$ Helsinki Institute of Physics, 
P.O. Box 9 (Siltavuorenpenger 20 C)\\
FIN-00014 University of Helsinki, Finland\\
${}^c$ Department of Physics, National Central University, Chungli, Taiwan 320,
ROC}

\maketitle

\begin{abstract}
Differential regularization
is used to investigate the one-loop quantum corrections to 
Chern-Simons-Maxwell spinor and scalar electrodynamics. We illustrate
the techniques to write the loop amplitudes in coordinate space.
The short-distance expansion method is developed to perform the 
Fourier transformation of the amplitudes into momentum space and the 
possible renormalization ambiguity in Chern-Simons type gauge theories 
in terms of differential regularization is discussed. We also stress
that the surface terms appearing in the differential regularization
should be kept along for finite theories and they will result in
the finite renormalization ambiguity.
\end{abstract}


\vspace{3ex}

 Differential regularization is a relatively new regularization scheme
$\cite{fjl}$. The basic idea of this regularization is quite simple. 
It works in coordinate space and is based on
the observation that the UV divergence reflects in 
the fact that the higher order amplitude
can not have a Fourier transform into momentum space due to the 
short-distance singularity. 
Thus one can regulate such an amplitude by first writing
its singular parts as the derivatives of the normal functions,
which have well defined Fourier transformation, and then by performing
the Fourier transformation  in partial integration and discarding 
the surface term, in this way one can directly get the renormalized result.
This regularization scheme successfully avoids the ambiguities of
the dimensional regularization 
in defining the dimensional continuation of ${\gamma}_5$-like objects 
since it does not need 
to continue the dimension of space-time. Up to now this method
has almost been verified in almost every field theory including 
the supersymmetric one$\cite{fjl,hl1,ha,hl2,mu}$. 
Its relation with the conventional
dimensional regularization and the compatibility with unitary
at two-loop level have also been investigated$\cite{dr,fjmv,sm}$. 
In some cases it indeed has advantages over all the conventional 
regularization schemes. 
Especially, it is very convenient to use this regularization
to discuss the conformal properties of quantum field theory
$\cite{os}$. 

In this letter we use this regularization to investigate the one-loop
quantum correction to Chern-Simons-Maxwell scalar and 
spinor electrodynamics\cite{djt}. One straightforward reason, 
as mentioned above, is that it avoids the ambiguity of 
dimensional continuation in 
defining the three-dimensional 
completely antisymmetric tensor ${\epsilon}_{\mu\nu\rho}$.  
As we know, the dimensionality in Chern-Simons-type theory
plays an important role since Chern-Simons
term is a topological one and the topological properties of theory
depend heavily on the three-dimensional antisymmetric
tensor, thus a calculation without using dimensional 
continuation is called for. The main motivation is that we want to
explore the possible origin of renormalization ambiguity of perturbative
Chern-Simons theory in the framework of dimensional regularization.
This ambiguity depends on the concrete regularization
schemes$\cite{coll,afll,lt}$ and  is the most puzzled feature 
of Chern-Simons type theories, up to now it has not been well
understood. Therefore, it is desirable to work in
a regularization scheme which does not greatly 
change the original theory.
Indeed, it has been found that higher covariant derivatived
Pauli-Villars regularization can bring non-physical quantum corrections
$\cite{mr}$, or at least this regularization
does not return back to the original theory when the regulator
is removed$\cite{af}$. We believe that up to now
differential regularization is the most appropriate method in this aspect
since it does not change the Lagrangian of the theory explicitly. 
Furthermore, from the view point of practical calculations, this
regularization is very suitable for the three-dimensional 
quantum field theory
since the propagators in three-dimensional space-time takes a
very simple form. In particular, the short-distance expansion
technique is developed in Ref.$\cite{clz}$, which can be used 
to calculate the one-loop quantum corrections exactly.

The Lagrangian in Euclidean space is as following 
\begin{eqnarray}
{\cal L}=-\frac{1}{4\lambda}F_{\mu\nu}F_{\mu\nu}-\frac{ik}{8\pi}
{\epsilon}_{\mu\nu\rho}A_{\mu}{\partial}_{\nu}A_{\rho}
-\frac{1}{2\alpha}({\partial}_{\mu}A_{\mu})^2-{\cal L}_{\rm matter}\, , 
\label{eq1}
\end{eqnarray}
where
\begin{eqnarray}
{\cal L}_{\rm matter}=D_{\mu}\phi^+D_{\mu}\phi+m^2\phi^+\phi
\end{eqnarray}
for scalar field and
\begin{eqnarray}
{\cal L}_{\rm matter}=\bar{\psi}({\gamma}_{\mu}D_{\mu}+im){\psi}
\end{eqnarray}
for the spinor case.
The ${\gamma}$ matrices  are defined as
\begin{eqnarray}
{\gamma}_{\mu}=i{\sigma}_{\mu},~~ {\gamma}_{\mu}{\gamma}_{\nu}
=-{\delta}_{\mu\nu}-{\epsilon}_{\mu\nu\rho}{\gamma}_{\rho},
~~\mbox{Tr}({\gamma}_{\mu}{\gamma}_{\nu})=-2{\delta}_{\mu\nu}\,. 
\label{eq2}
\end{eqnarray}
The propagators for electron, scalar field and gauge field take very
simple forms in coordinate space:
\addtocounter{equation}{1}
$$
S(x)=({\gamma}_{\mu}{\partial}_{\mu}-im)\frac{1}{4\pi}\frac{1}{x}
e^{-mx}\,, 
\label{eq3}\eqno{(5a)}$$
$$D(x)=\frac{1}{4\pi}\frac{1}{x}e^{-mx}, \eqno{(5b)} $$
$$G_{\mu\nu}(x)=\left[\frac{i}{k/(4\pi)}{\epsilon}_{\mu\nu\rho}{\partial}_{\rho}
-\frac{\lambda}{n^2}(
{\partial}^2{\delta}_{\mu\nu}-{\partial}_{\mu}{\partial}_{\nu})\right]
\frac{1}{4\pi}\frac{1}{x}\left(1-e^{-nx}\right)\, , 
\label{eq31}\eqno{(5c)}$$
where (and in what follows)
$x{\equiv}|x|$, $n{\equiv}\displaystyle\frac{{\lambda}k}{4\pi}$ 
and we work in Lauge gauge (${\alpha}=0$).

Now we calculate the vacuum polarization tensor. Let us first see the
contribution from electron loop:
\begin{eqnarray}
{\Pi}^{\rm (spinor)}_{\mu\nu}(x)&=&
-\mbox{Tr}[{\Gamma}_{\mu}S(x){\Gamma}_{\nu}S(-x)]
\nonumber\\[2mm]
&=&-\frac{e^2}{8{\pi}^2}\left[2x_{\mu}x_{\nu}\left(
\frac{1}{x^6}+\frac{2m}{x^5}+\frac{m^2}{x^4}\right)e^{-2mx}
\right.\nonumber\\[2mm]
&-&{\delta}_{\mu\nu}\left(\frac{1}{x^4}+
\frac{2m}{x^3}+\frac{2m^2}{x^2}\right)e^{-2mx}\nonumber
\\[2mm]
&-&\left. 2im{\epsilon}_{\mu\nu\rho}{\partial}_{\rho}
\left(\frac{1}{x}e^{-mx}\right)
\frac{1}{x}e^{-mx}\right]\, .
\label{eq8}
\end{eqnarray}
Obviously, the terms $1/x^n$ with $n{\geq}3$ can not
perform their Fourier transformation into momentum space. According
to the idea of differential regularization, 
the vacuum polarization tensor can be written as the differential 
regulated version 
\begin{eqnarray}
{\Pi}^{\rm (spinor)}_{\mu\nu}(x)&=&-\frac{e^2}{8\pi ^2}({\partial}_{\mu}{\partial}_{\nu}
-{\delta}_{\mu\nu}{\partial}^2)\left[\left(\frac{1}{4}\frac{1}{x^2}  
+\frac{1}{2}\frac{m}{x}\right)e^{-2mx}+m^2\mbox{Ei}(-2mx)\right]\nonumber
\\[2mm]
&+&\frac{ie^2}{8{\pi}^2}m{\epsilon}_{\mu\nu\rho}{\partial}_{\rho}
\left(\frac{1}{x^2}e^{-2mx}\right)\, 
\label{eq9}
\end{eqnarray}
with $\mbox{Ei}(y)=\displaystyle {\int}_y^{\infty}e^{-t}t^{-1}dt$ being
the exponential integral function.
One can see that there is no new dimensional 
parameter appearing in Eq.(\ref{eq9}), which means the
finiteness of the vacuum polarization tensor. As  
suggested in Ref.\cite{fjl} and developed in Ref.\cite{clz}, we use the
short-distance technique to perform Fourier transformation into momentum space 
so that we can preserve the possible nonvanishing surface term.
With aid of the formulas
\begin{eqnarray}
{\int}_{R^3_{\epsilon}}d^3x{\partial}_{\mu}f(x)e^{ip{\cdot}x}
&=&4{\pi}i\frac{\partial}{{\partial}p_{\mu}}
\left[\frac{\sin(p\epsilon)}{p}\right]f({\epsilon})
-ip_{\mu}{\int}_{R^3_{\epsilon}}d^3 xe^{ip{\cdot}x}f(x)\, ,
\nonumber\\[2mm]
{\int}_{R^3_{\epsilon}}e^{ip{\cdot}x}{\partial}_{\mu}{\partial}_{\nu}
f(x)&=&4{\pi}\frac{{\partial}^2}{{\partial}p_{\mu}{\partial}p_{\nu}}
\left[\frac{\sin(p\epsilon)}{p\epsilon}\right]
\frac{d}{dx}f(x)|_{x=\epsilon}
-4{\pi}p_{\mu}\frac{\partial}{{\partial}p_{\nu}}
\left[\frac{\sin(p\epsilon)}{p}\right]f(\epsilon)
\nonumber\\[2mm]
&-&p_{\mu}p_{\nu}{\int}_{R^3_{\epsilon}}d^3xf(x)e^{ip{\cdot}x}\,,
\label{eq10}
\end{eqnarray}
where $p{\equiv}|p|$, $R^3_{\epsilon}$ denotes the integration region
$R^3-B_{\epsilon}$ and $B_{\epsilon}$ is a small sphere of
radius $\epsilon$ around the origin, we obtain 
\begin{eqnarray}
{\Pi}^{\rm (spinor)}_{\mu\nu}(p)
&=&\frac{e^2}{2\pi}\left\{{\epsilon}_{\mu\nu\rho}
p_{\rho}\frac{m}{p}\arctan\frac{p}{2m}-
(p^2{\delta}_{\mu\nu}-p_{\mu}p_{\nu})\left[\frac{m}{2p^2}
\right.\right.\nonumber
\\[2mm]
&+& \left.\left.\frac{1}{4}\left(\frac{m}{p}-\frac{4m^2}{p^3}\right)
\arctan\frac{p}{2m}\right]\right\}\, .
\label{eq11}
\end{eqnarray} 
In deriving (\ref{eq11}) we have taken the limit ${\epsilon}{\rightarrow}0$ 
after performing the integration.
The necessity of preserving the
surface term should be stressed for a finite theory. Otherwise, if one
throws away the surface term (as that suggested in the original
paper $\cite{fjl}$ on differential regularization), some finite terms
as $\epsilon{\rightarrow}0$ will lose, this will certainly lead 
to the ambiguity of finite renormalization.

The vacuum polarization tensor in scalar case 
is a little complicated since
there is derivative on the vertex, but the calculation is straightforward.
It reads as 
\begin{eqnarray}
\Pi^{\rm (scalar)}_{\mu\nu}(x)&=&e^2[{\partial}_{\mu}D(-x){\partial}_{\nu}D(x)
-D(x){\partial}_{\mu}{\partial}_{\nu}D(-x)
-D(-x){\partial}_{\mu}
{\partial}_{\nu}D(x)+{\partial}_{\mu}D(x){\partial}_{\nu}D(-x)]
\nonumber\\[2mm] 
&=&\frac{e^2}{8{\pi}^2}\left\{\frac{x_{\mu}x_{\nu}}{x^2}
\left[\frac{d}{dx}D(x)\right]^2-\frac{x_{\mu}x_{\nu}}{x}D(x)\frac{d}{dx}
\left[\frac{1}{x}\frac{d}{dx}D(x)\right]-D(x)\frac{{\delta}_{\mu\nu}}{x}
\frac{d}{dx}D(x)\right\}\nonumber\\[2mm]
&=&\frac{e^2}{8{\pi}^2}\left[{\delta}_{\mu\nu}
\left(\frac{1}{x^4}+\frac{m}{x^3}\right)e^{-2mx}-x_{\mu}x_{\nu}
\left(\frac{2}{x^6}+\frac{m}{x^5}\right)e^{-2mx}\right]\nonumber\\[2mm]
&=&\frac{e^2}{16\pi^2}({\partial}_{\mu}{\partial}_{\nu}-{\delta}_{\mu\nu}
{\partial}^2)\left[-\frac{1}{2x^2}e^{-2mx}+\frac{m}{x}
e^{-2mx}+2m^2\mbox{Ei}(-2mx)\right]\, .
\end{eqnarray}
Its Fourier transformation is read as
\begin{eqnarray}
{\Pi}^{\rm (scalar)}_{\mu\nu}(p)=\frac{e^2}{4\pi}(p^2g_{\mu\nu}-p_{\mu}p_{\nu})
\left[\frac{m}{p^2}-\frac{1}{2p}\arctan\frac{p}{m}-
\frac{2m^2}{p^3}\arctan\frac{p}{2m}\right]\, .
\end{eqnarray}

The other one-loop two-point functions are self-energy of matter fields.
The amplitude for the self-energy of electron is 
\begin{eqnarray}
{\Sigma}(x)&=&{\Gamma}_{\nu}S(-x){\Gamma}_{\mu}(x)G_{\mu\nu}(x)
\nonumber\\[2mm]
&=&\frac{e^2}{16{\pi}^2}\left\{
\frac{2i}{k/(4\pi)}\left[\left(\frac{1}{x^4}+\frac{m}{x^3}\right)
e^{-mx}-\left(\frac{1}{x^4}+\frac{m+n}{x^3}
+\frac{mn}{x^2}\right)e^{-(m+n)x}\right]\right.
\nonumber\\[2mm]
&-&\frac{2m}{k/(4\pi)}{\gamma}_{\rho}x_{\rho}\left[-\frac{1}{x^4}
e^{-mx}+\left(\frac{1}{x^4}+\frac{n}{x^3}\right)e^{-(m+n)x}\right]
+\frac{\lambda}{n^2}{\gamma}_{\mu}x_{\mu}\left[\left(\frac{4}{x^6}+
\frac{4m}{x^5}\right)e^{-mx}\right.\nonumber\\[2mm]
&+&\left(-\frac{4}{x^6}-\frac{4(m+n)}{x^5}+
\frac{-4mn-2n^2}{x^4}-\frac{2mn^2}{x^3}\right)e^{-(m+n)x}
\nonumber\\[2mm]
&-&\left.\left.\frac{2im{\lambda}}{x^2}e^{-(m+n)x}\right]\right\}\, .
\label{eq13}
\end{eqnarray} 
With similar operations as above, we write (\ref{eq13}) as the differential
regulated form
\begin{eqnarray}
{\Sigma}(x)&=&\frac{e^2}{16{\pi}^2}\left\{{\gamma}_{\mu}{\partial}_{\mu}
\left[-\frac{2m}{k/(4\pi)}\left((\frac{1}{2x^2}-\frac{m}{2x})e^{-mx}+
(\frac{1}{2x^2}+\frac{m-n}{2x})e^{-(m+n)x}\right.\right.
\right.\nonumber\\[2mm]
&-&\frac{m^2}{2}\mbox{Ei}(-mx)
+\left.\frac{m^2-n^2}{2}\mbox{Ei}[-(m+n)x]\right)
-{\lambda}\left((-\frac{1}{2x^2}-\frac{m-n}{2x})e^{-(m+n)x}
\right.\nonumber\\[2mm]
&+&\left.\left.
\frac{n^2-m^2}{2}\mbox{Ei}[-(m+n)x]\right)\right]
+\frac{\lambda}{n^2}\left[-{\partial}^2[\frac{1}{2}\frac{1}{x^2}
e^{-mx}(1-e^{-nx})]+\left(
\frac{m^2}{x^2}-\frac{m^3}{2x}\right)e^{-mx}\right.
\nonumber\\[2mm]
&+&\left(\frac{2m^2-n^2}{2x^2} 
-\frac{m^2(m+n)}{2x}\right)e^{-(m+n)x}-m^4\mbox{Ei}(-mx)
\nonumber\\[2mm] 
&+&
\left.(\frac{m^4}{2}-\frac{3m^2n^2}{8}-\frac{n^4}{4})\mbox{Ei}[-(m+n)x]\right]
+\frac{2i}{k/(4\pi)}\left[{\partial}^2[\frac{1}{2x^2}
e^{-mx}(1-e^{-nx})]\right.\nonumber\\[2mm]
&-&\left.\left.\frac{m^2}{2x^2}e^{-mx}+
\frac{m^2+n^2}{2x^2}e^{-(m+n)x}\right]-
\frac{2im{\lambda}}{x^2}e^{-(m+n)x}\right\}\, .
\end{eqnarray}
Using the short-distance expansion (\ref{eq10}), we obtain the
electron self-energy in momentum space as
\begin{eqnarray}
{\Sigma}(p)&=&\frac{e^2}{4\pi}i{\gamma}_{\mu}p_{\mu}
\left\{\frac{2m}{k/(4\pi)}
\left[\frac{n}{2p}+\left(\frac{1}{2}+\frac{m^2}{2p^2}\right)
\arctan\frac{p}{m}-\left(\frac{1}{2}+\frac{m^2-n^2}{2p^2}
\right)\arctan\frac{p}{m+n}\right.
\right.\nonumber\\[2mm]
&-&\lambda\left[\frac{m-n}{2p}+\left(\frac{1}{2}-\frac{m^2-n^2}{2p^2}\right)
\arctan\frac{p}{m+n}\right]+
\frac{\lambda}{n^2}\left[\frac{n}{2}+\frac{m^3}{2p^2}\right.\nonumber\\[2mm]
&-&(\frac{p}{2}+\frac{m^2}{p^2}-
\frac{m^4}{2p^3})\arctan\frac{p}{m}
+(\frac{p}{2}+\frac{m^2}{p}+
\frac{n^2}{2p}-\frac{m^4}{2p^3}+\frac{3}{8}\frac{m^2n^2}{p^3}+
\frac{n^4}{4p^3})\arctan\frac{p}{m+n}\nonumber\\[2mm]
&+&\left.\left.
(\frac{m^4}{2}-\frac{3m^2n^2}{8}-\frac{n^4}{4})\frac{m+n}{p^2[p^2+(m+n)^2]}
-\frac{m^2(m-n)}{2[p^2+(m+n)^2]}\right]\right\}
\nonumber\\[2mm]
&+&\frac{e^2}{4\pi}\left\{\frac{2i}{k/(4\pi)}\left[\frac{n}{2}-
(\frac{p}{2}+\frac{m^2}{2p})\arctan\frac{p}{m}
-\frac{m^2-n^2}{2p}\arctan\frac{p}{m+n}\right]\right.\nonumber\\[2mm]
&-&\left.\frac{2im{\lambda}}{p}\arctan\frac{p}{m+n}\right\}\,.
\end{eqnarray}

As for the self-energy of the scalar field, since there are derivatives 
in the interaction vertex, usually it is very confusing to decide the 
derivative acting on external or internal lines when writing the amplitude 
in the coordinate space. The key technique is 
directly writing the amplitude in the Fourier transformed form and this
can clearly show the action of the derivative in the vertex on external
legs or internal lines, 
\begin{eqnarray}
{\Omega}(p)&=&e^2{\int}d^3xe^{ip{\cdot}x}\left[ip_{\mu}{\partial}_{\nu}
D(x)G_{\mu\nu}(x)+p_{\mu}p_{\nu}D(x)G_{\mu\nu}(x)\right.\nonumber\\[2mm]
&-&\left.{\partial}_{\mu}{\partial}_{\nu}
D(x)G_{\mu\nu}(x)+ip_{\nu}{\partial}_{\mu}
D(x)G_{\mu\nu}(x)\right]\,. 
\label{eq:se}  
\end{eqnarray}
Expanding each term in Eq.(\ref{eq:se}) and writing them
in the derivative form, we obtain the scalar self-energy 
\begin{eqnarray}
{\Omega}(p)&=&\frac{e^2}{16{\pi}^2}\frac{\lambda}{n^2}
{\int}d^3xe^{ip{\cdot}x}ip_{\mu}{\partial}_{\mu}
\left\{
{\partial}^2\left[\frac{1}{2x^2}\left(e^{-mx}-e^{-(m+n)x}\right)
\right]\right.\nonumber\\[2mm]
&+&\left[\frac{m^2+n^2}{x^2}-\frac{(m-n)^2(m+n)}{2x}
\right]e^{-(m+n)x}\nonumber\\[2mm]
&+&\left.\frac{m^3}{2x}e^{-mx}
+\frac{m^4}{2}\mbox{Ei}(-mx)-\frac{(m^2-n^2)^2}{2}\mbox{Ei}[-(m+n)x]\right\}
\nonumber\\[2mm]
&+&\frac{e^2}{16{\pi}^2}\frac{\lambda}{n^2}
{\int}d^3xe^{ip{\cdot}x}p_{\mu}p_{\nu}\left\{
{\partial}_{\mu}{\partial}_{\nu}\left[\left(-\frac{m^2}{16}+
\frac{3}{8x^2}-\frac{5m}{8x}+\frac{m^3x}{16}\right)e^{-mx}
\right.\right.\nonumber\\[2mm]
&+&\left(\frac{(m-n)^2}{16}-\frac{3}{8x^2}+\frac{5m-3n}{8x}
-\frac{x(m+n)(m-n)^2}{16}\right)e^{-(m+n)x}\nonumber\\[2mm]
&+&\left.\left(-\frac{3m^2}{4}+\frac{m^4x^2}{16}\right)\mbox{Ei}(-mx)+
\left(\frac{3m^2-n^2}{4}-\frac{(m^2-n^2)^2x^2}{16}\right)\mbox{Ei}[-
(m+n)x]\right]
\nonumber\\[2mm]
&+&{\delta}_{\mu\nu}\left[{\partial}^2\left(-\frac{1}{8x^2}
e^{-mx}+\frac{1}{8x^2}e^{-(m+n)x}\right)
+\left(\frac{m^2}{4x^2}-\frac{m^3}{8x}\right)e^{-mx}\right.\nonumber\\[2mm]
&+&\left(
\frac{3n^2-m^2}{4x^2}+\frac{(m-n)^2(m+n)}{8x}\right)e^{-(m+n)x}
\nonumber\\[2mm]
&-&\left.\left.\frac{m^4}{8}\mbox{Ei}(-mx)+\frac{(m^2-n^2)^2}{8}\mbox{Ei}[-(m+n)x]\right]
\right\}\nonumber\\[2mm]
&+&\frac{e^2}{16{\pi}^2}\frac{\lambda}{n^2}{\int}d^3xe^{ip{\cdot}x}
\left\{{\partial}^2\left[
{\partial}^2\left(-\frac{1}{4x^2}e^{-mx}+\frac{1}{4x^2}
e^{-(m+n)x}\right)
+\frac{m^2}{2x^2}e^{-mx}\right.\right.
\nonumber\\[2mm]
&-&\left.\left.\frac{m^2+n^2}{2x^2}
e^{-(m+n)x}\right]-\frac{m^2n^2}{2x^2}e^{-(m+n)x}\right\} \, .
\end{eqnarray}
After performing Fourier transformation through the short distance
expansion, we have
\begin{eqnarray}
{\Omega}(p)&=&\frac{e^2}{4\pi}\frac{\lambda}{n^2}\left\{\frac{mn(m+n)}{2}
-\frac{7n^3}{6}+n p^2-\frac{9m^5}{8(p^2+m^2)}+\frac{5m^7}{4(p^2+m^2)^2}
-\frac{m^9}{2(p^2+m^2)^3}\right.\nonumber\\[2mm]
&+&\frac{9(m-n)^2(m+n)^3}{8[p^2+(m+n)^2]}-
\frac{5(m-n)^2(m+n)^5}{4[p^2+(m+n)^2]^2}
+\frac{(m-n)^2(m+n)^7}{2[p^2+(m+n)^2]^3}\nonumber
\\[2mm]
&+&\left[-\frac{3m^4}{8p}
-\frac{p^3}{2}-\frac{3m^4p}{p^2+m^2}+\frac{6m^6p}{(p^2+m^2)^2}-
\frac{4m^8p}{(p^2+m^2)^3}\right]\arctan\frac{p}{m}
+ \left[\frac{p^3}{2} \right.\nonumber
\\[2mm]
&+&\frac{[(m-n)^2+5(m^2+n^2)]p}{8}+
\frac{(m^2-n^2)^2+2(m^4+n^4)}{8p}+\frac{3(m^2-n^2)^2p}{8[p^2+(m+n)^2]}
\nonumber\\[2mm]
&-&\left.\left.\frac{3(m-n)^2(m+n)^4p}{4[p^2+(m+n)^2]^2}+
\frac{(m-n)^2(m+n)^6p}{2[p^2+(m+n)^2]^3}\right]\arctan\frac{p}{m+n}
\right\}\, .
\end{eqnarray}

Since the complete one-loop amplitude is already given, the finite
renormalization can be easily performed by choosing a renormalization point
(for example, one typical choice is $p=0$), as usual, we can define
various renormalization constants and the radiative corrections.

In summary, we have shown how differential regularization can be 
used to investigate the one-loop two-point functions of 
three-dimensional Chern-Simons-Maxwell 
spinor and scalar electrodynamics. For the scalar case, 
where there is derivative on the vertex, we develop the technique to 
distinguish how the derivatives act on an external or an 
internal line properly when writing the amplitude in coordinate space. 
In particular, using the short-distance 
expansion technique, we show how a renormalization ambiguity can be
generated for a finite theory.
For example, in the Fourier transform of the 
term ${\partial}^2[(1-e^{-mx})/x^2]$,
if we consider the surface term through the short-distance expansion
\begin{eqnarray}
{\int}_{R^3_\epsilon}d^3x{\partial}^2\left(\frac{1-e^{-mx}}{x^2}\right)
e^{ip{\cdot}x}
=4\pi\left[m-\frac{{\pi}p}{2}+p\arctan\frac{p}{m}\right]\, ,
\end{eqnarray}
a non-vanishing surface term $4{\pi}m$ appears. However, if we perform
the Fourier transform according to the original idea of differential
regularization$\cite{fjl}$, this surface term is usually discarded. 
So we see that the surface terms are very important 
in Chern-Simons type theories since they are usually 
finite$\cite{bc,dpls,bms}$ at least at the one-loop 
level$\cite{sp}$, and thus the non-vanishing surface term would result 
in an ambiguity in defining the finite renormalization.  As we know, 
in a finite theory, the $\beta$-function and anomalous dimensions 
of each field vanish, the renormalization group equation
is trivial and the only criterion for the equivalence among
different renormalization conditions is that all the regularization  
schemes preserving the fundamental symmetry such as gauge invariance 
should give the same gauge invariant radiative corrections, 
so that the finite renormalization ambiguity is a serious problem. 
Therefore, we believe that in this respect differential regularization, 
thanks to its nature, provides with a better understanding of the 
renormalization ambiguity in Chern-Simons type theories.
 
\acknowledgments
The support of the Academy of Finland is greatly acknowledged. W.F.C. thanks
the World Laboratory, Switzerland, for financial
support. We thank M. Asorey and  F. Ruiz Ruiz for their useful
discussions on Chern-Simons theory and especially
Z.Y. Zhu for his help in understanding the differential regularization.
H.C.L. is partially supported by grant 85-2112-M-008-011 from the National
Science Council, ROC.


\begin{thebibliography}{99}

\bibitem{fjl}  D.Z. Freedman, K. Johnson and J.I. Latorre, 
Nucl. Phys. {\bf B371} (1992) 353.

\bibitem{hl1}  P.E. Haagensen and J.I. Latorre, Phys. Lett. {\bf B283}
(1992) 293.

\bibitem{ha}  P.E. Haagensen, Mod. Phys. Lett. {\bf A7} (1992) 893.

\bibitem{hl2}  P.E. Haagensen and J.I. Latorre, Ann. Phys. {\bf 221} (1993)
77.

\bibitem{mu}  R. Munoz-Tapia, Phys. Lett. {\bf B295} (1992) 95.

\bibitem{dr}  G. Dunne and N. Rius, Phys. lett. {\bf B293} (1992) 367.

\bibitem{fjmv}  D.Z. Freedman, K. Johnson, R. Munoz-Tapia and X.
Vilasis-Cardona, Nucl. Phys. {\bf B395} (1993) 454.

\bibitem{sm}  V.A. Smirnov, Nucl. Phys. {\bf B427} (1994) 325; Z. Phys.
{\bf C67} (1995) 531.

\bibitem{os} J. Erdmenger and H. Osborn, Nucl. Phys. {\bf B483} (1997) 431.

\bibitem{djt}  S. Deser, R. Jackiw and S. Templeton, Phys. Rev. Lett. 
{\bf 48} (1982) 975; Phys. Rev. {\bf D23} (1981) 2291; 
Ann. Phys. {\bf 140} (1982) 372.

\bibitem{coll} E. Guadagnini, M. Matellini and M. Mintchev,
Phys. Lett. {\bf B227} (1989) 111; 
W. Chen, G. W. Semenoff and Y.S. Wu, Mod. Phys. Lett. {\bf A5}
(1990) 1833; Phys. Rev. {\bf D46} (1992) 5521; L. Alvarez-Gaum\'{e}, J.M.F.
Labastida and A.F. Ramallo, Nucl. Phys. {\bf B334} (1990) 103; M. Asorey
and F. Falceto, Phys. Lett. {\bf B241} (1990) 31; C. P. Martin, Phys.
Lett. {\bf B241} (1990) 513; G. Giavarini, C.P. Martin and F. Ruiz Ruiz,
Nucl. Phys. {\bf B381} (1992) 222, Phys. Rev. {\bf D47} (1993) 5536;
W. F. Chen and Z.Y. Zhu, J. Phys. {\bf A27} (1994) 1781.

\bibitem{afll} M. Asorey, F. Falceto, J. L. L\'{o}pez and G. Luz\'{o}n,
Phys. Rev. {\bf D49} (1994) 5377; Nucl. Phys. {\bf B429} (1994) 344.  

\bibitem{lt} M. Leblanc and M.T. Thomaz, Phys. Lett. {\bf B281} (1992)
259.

\bibitem{mr} C. P. Martin and F. Ruiz Ruiz, Nucl. Phys. {\bf B436}
(1995) 545.

\bibitem{af} M. Asorey and F. Falceto, Phys. Rev. {\bf D54} (1996) 
5290.

\bibitem{clz} W.F. Chen, H.C. Lee and Z.Y. Zhu, Phys. Rev. {\bf D55} (1997)
3664.

\bibitem{bc} A. Blasi and R. Collina, Nucl. Phys. {\bf B345} (1990) 472.

\bibitem{dpls} F. Delduc, O. Piguet, C. Lucchesi and S.P. Sorella, 
Nucl. Phys. {\bf B346} (1990) 313.

\bibitem{bms} A. Blasi, N. Maggiore and S.P. Sorella, Phys. Lett. {\bf 
B285} (1992) 54.

\bibitem{sp} E.R. Speer, J. Math. Phys. {\bf 15} (1974) 1.


\end{thebibliography}
\end{document}